# Double-Layered Silica-Engineered Fluorescent Nanodiamonds for Catalytic Generation and Quantum Sensing of Active Radicals

Jia Su[1,2]†, Zenghao Kong[1,3]†, Fei Kong[1,2,3]†, Xing Liu[4]†, Linyu Zeng[1,2], Zhecheng Wang[1,2,5], Zijian Zeng[1], Jie Liu[4], Jihu Su[1,2,3], Junhua Yuan[1,2], Guosheng Shi[4*], Fazhan Shi[1,2,3,5,6*]

[1]Laboratory of Spin Magnetic Resonance, School of Physical Sciences, Anhui Province Key Laboratory of Scientific Instrument Development and Application, University of Science and Technology of China, Hefei 230026, China

[2]Hefei National Research Center for Physical Sciences at the Microscale, University of Science and Technology of China, Hefei 230026, China

[3]Hefei National Laboratory, University of Science and Technology of China, Hefei 230088, China

[4]Shanghai Applied Radiation Institute, State Key Lab. Advanced Special Steel, Shanghai University, Shanghai 200444, China.

[5]School of Biomedical Engineering and Suzhou Institute for Advanced Research, University of Science and Technology of China, Suzhou 215123, China

[6]The First Affiliated Hospital of USTC, Division of Life Sciences and Medicine, University of Science and Technology of China, Hefei 230026, China

*Corresponding author(s): Guosheng Shi and Fazhan Shi

Email: gsshi@shu.edu.cn; fzshi@ustc.edu.cn.

†These authors contributed equally to this work.

**Abstract**

Fluorescent nanodiamonds (FNDs) hosting nitrogen-vacancy (NV) centers have attracted considerable attention for quantum sensing applications, particularly owing to notable advancements achieved in the field of weak magnetic signal detection in recent years. Here, we report a practical quantum-sensing platform for the controlled production and real-time monitoring of ultra-short-lived reactive free radicals using a double-layered silica modification strategy. An inner dense silica layer preserves the intrinsic properties of NV centers, while an outer porous silica layer facilitates efficient adsorption and stabilization of hydroxyl radicals and their precursor reactants. By doping this mesoporous shell with gadolinium (III) catalysts, we achieve sustained, light-free generation of hydroxyl radicals via catalytic water splitting, eliminating reliance on external precursors. The mechanism underlying this efficient radical generation is discussed in detail. The radical production is monitored in real time and in situ through spin-dependent $T_1$ relaxometry of the NV centers, demonstrating stable and tunable radical fluxes, with concentration tunable across a continuous range from approximately 100 mM to molar levels by adjusting the catalyst condition. This study extends the technical application of nanodiamonds from relaxation sensing to the controlled synthesis of reactive free radicals, thereby providing robust experimental evidence to support the advancement of quantum sensing systems in intelligent manufacturing.

**Significance Statement**

This work establishes a practical quantum-sensing platform for the controlled generation and real-time detection of hydroxyl radicals—a highly reactive and biologically critical species that has been exceptionally difficult to monitor directly. By integrating a protective yet accessible silica interface with embedded catalysts, we enable continuous hydroxyl radical production via water splitting and its immediate readout using nitrogen-vacancy centers in nanodiamonds. This approach removes the reliance on external photo-activation of semiconductor nanoparticles or chemical precursors such as hydrogen peroxide, offering a stable, tunable, and minimally invasive method to transient radical production and study transient radical dynamics. It bridges quantum sensing with catalytic chemistry, opening avenues for real-time, in-situ investigation of active radicals and oxidative processes across diverse fields.

**Main Text**

**Introduction**

Fluorescent nanodiamonds (FNDs)[1] hosting nitrogen-vacancy (NV) centers[2] have emerged as a powerful platform for quantum sensing and imaging[3,4,5]. These atomic-scale defects, consisting of a nitrogen atom adjacent to a lattice vacancy, possess optically addressable electron spin states ($S = 1$) that enable precision measurements of magnetic fields[6,7] and temperature[8,9]. FND-based relaxometry[10] has demonstrated high sensitivity toward paramagnetic ions[11,12,13,14] and enables real-time monitoring of localized chemical changes, including pH variations[15], ascorbic acid oxidation[16] and hydrogen peroxide decomposition[12]. This technique further facilitates continuous observation of radical dynamics, offering distinct advantages in detecting highly reactive species such as hydroxyl[17,18] and nitroxide radicals[19]—key intermediates in cellular metabolism[20,21] and physiological regulation. Conventional organic fluorescent molecular sensors for free radicals sensing operate through radical coupling reactions, yet they are inevitably plagued by the insidious accumulation of signals over time—a flaw that undermines their fidelity. A more detailed comparison of various methods for the detection is provided in Supplementary Table1. In contrast, FNDs neither degrade with use nor retain echoes of past signals. This exceptional durability enables FNDs to provide not only real-time but also highly reliable monitoring of dynamic processes

Among reactive oxygen species, the hydroxyl radical ($\cdot OH$) is of paramount interest due to its extreme reactivity, characterized by a one-electron reduction potential of 2.31 V[22] and rapid reaction kinetics ($10^9$–$10^{10}$ $M^{-1} \cdot s^{-1}$)[23], and central role in oxidative stress. However, its ultra-short lifetime (~$10^{-9}$ s)[23] and the limitations of conventional peroxide-dependent generation methods pose significant challenges for its controlled production and real-time detection (A comprehensive comparison of more hydroxyl radical detection and application strategies is provided in Supplementary Tables 1 and 2.). Achieving real-time and in-situ detection of reactive oxygen species is not only of great significance for understanding complex chemical and biological processes (such as catalytic reaction kinetics or cellular oxidative stress responses), but also offers the possibility of dynamic regulation of synthetic pathways. By feeding back sensing signals to the reaction system, it is expected to establish a closed-loop controlled synthetic platform, precisely adjusting reaction conditions and parameters, thereby enabling selective control over the generation rate for various applications.

Despite the potential of FNDs as a highly sensitive paramagnetic sensing platform, their practical application in the detection of reactive oxidative species confronts two fundamental challenges: Firstly, constructing an interface capable of efficiently capturing and stabilizing short-lived reactive species near the sensor surface is of particular criticality. Such reactive species—including hydroxyl radicals—exhibit extremely high chemical reactivity and exceptionally short half-lives, typically undergoing fast reaction or decomposition. An ideal sensing interface must not only incorporate a high-affinity functionalized layer to enrich target species but also extend their residence time in the sensing region, thereby enhancing detection probability and signal-to-noise ratio. Concurrently, this interface design must preclude the introduction of any non-target factors that might disrupt the electronic spin states of NV centers, so as to preserve their quantum relaxation properties and high-sensitivity detection capabilities. Secondly, NV centers located near the diamond surface are highly susceptible to external environmental perturbations—particularly magnetic noise which significantly degrade the stability and resolution of their quantum readout. To address this, effective surface engineering strategies[24,25] must be implemented to suppress surface state density and enhance structural robustness. Notably, however, excessive surface passivation can increase transfer resistance, impeding the diffusion of target analytes to the sensing region and thereby compromising response speed and sensitivity. Consequently, achieving an optimal balance between surface protection and analyte accessibility has emerged as a core challenge in the development of high-performance FND sensors.

Here, we address these challenges by developing a rationally designed nanoplatform that integrates sustained and stable hydroxyl radical generation with real-time in situ quantum detection. Our approach features a double-layered silica architecture on FNDs (Figure 1): a dense inner shell preserves the NV centers' quantum coherence, while a porous outer layer serves as a high-surface-area scaffold for radical stabilization. By incorporating metal-ion catalysts into this shell, continuous hydroxyl radical generation is achieved through water splitting, thereby eliminating reliance on external photoirradiation or peroxide-based precursors. Given that water molecules typically exhibit limited spontaneous dissociation into hydrogen and hydroxyl radicals, the underlying mechanism enabling efficient radical production is thoroughly examined. The hydroxyl radical generation is directly monitored in real time through spin-dependent NV center readout, demonstrating stable and catalyst-tunable radical fluxes. This work establishes a robust platform for on-demand hydroxyl radical generation and sensing, paving the way for persistent, in-situ monitoring of transient species and redox dynamics in complex biological and chemical environments. It further advances the integration of materials science and quantum technology, offering new opportunities for fields such as intelligent manufacturing.

**Results**

**Synthesis and characterization of MS-silica-FND**

The mesoporous silica (MS) coating was synthesized using the Stöber method[26], employing micelles formed from a cationic surfactant as the template[27]. Prior to applying the MS coating to the surface of FNDs, a thin silica layer was first deposited on the FND surface (silica-FND) to improve the relaxation properties of NV centers by reducing magnetic noise from the surface. Transmission electron microscopy (TEM) images (Figure 2a) reveal that the prepared silica-FND exhibits a core-shell structure, with the silica shell approximately 9.4 ± 2.1 nm thick (statistical data shown in Figure 2b). The mesoporous silica-coated nanodiamonds (MS-silica-FND) displayed a spherical morphology, with a diameter of approximately 127.3 ± 20.8 nm. By controlling the synthesis conditions, the thickness of the silica layer was varied between 2.0 ± 0.8 nm and 12.4 ± 2.1 nm (Supplementary Figure 1-2). Dynamic light scattering (DLS) data (Figure 2c) further confirmed the successful functionalization of the nanodiamond, other physicochemical properties of the core-shell functionalized nanodiamonds were systematically characterized (Supplementary Figure 3-4). Nanodiamonds containing NV centers at the core of MS-silica-FND are designed to directly monitor the local concentration of paramagnetic hydroxyl radicals. The electronic structure of the negatively charged NV center (electron spin S = 1) allows initialization and readout of spin-sublevels by optical means. A simplified energy diagram of NV$^-$ centers in diamond is shown in Figure 2d. The identifying features of NV$^-$ is zero field magnetic resonance at ~2.87 GHz. This magnetic resonance occurs between the $|m_s = 0\rangle$ and $|m_s = ±1\rangle$ and can be detected by optically detected magnetic resonance (ODMR). The sensing technique FND relaxometry utilized in this study is based on $T_1$ relaxation measurements. Specifically, the electron in the NV center is polarized into the $|m_s = 0\rangle$ state of the ground state under continuous illumination by a 532 nm laser. Upon cessation of the laser, the electron progressively relaxes to the $|m_s = ±1\rangle$ states of the ground state, ultimately reaching thermal equilibrium. In the presence of high-frequency magnetic noise in the environment, the time required to reach thermal equilibrium may be prolonged. For a single NV center, the decay process follows a mono-exponential pattern, described by the equation: $I(\tau) = I_0(1 + Ce^{-\tau/T_1})$, where τ is the duration time that the NV spin remains in the dark after being initialized into the $|m_s = 0\rangle$ spin sublevel with an optical pulse and the characteristic relaxation time is denoted as $T_1$[11,28]. Statistical single-particle measurements reveal a relatively broad distribution of $T_1$ times across individual particles (Figure 2e). The differences among individual FNDs have attracted extensive attention and widely known[29]. Variations in the number, positioning, and properties of NV centers can be significant among different FNDs. Measurements of large ensembles show that the ensemble-averaged results are in good agreement with the single-particle statistical measurements and exhibit high reproducibility (Figure 2f). To maintain high sensitivity while enhancing the reliability and reproducibility of the results, this study combines single-particle analysis with ensemble studies.

**Highly reactive radical generation via light-free catalytic water splitting**

Figure 3a displays the time profiles of changed relaxation rates for NV centers in MS-silica-FND immersed in a 1 µM Gadolinium (III) diethylenetriaminepentaacetate dianion (Gd(DTPA)$^{2-}$) solution. The increased changed relaxation rate ($\Delta\Gamma = \frac{1}{T_1} - \frac{1}{T_1^{(0)}}$) is proportional to the number of magnetic species, indicating a rise in the concentration of paramagnetic species surrounding MS-silica-FND. However, silica-FND at the same gadolinium (III) concentration (1 µM) with no significant change was observed. The stability of the silica-FND with respect to their altered relaxation rates reflects multiple critical aspects. First, it demonstrates that surface modification with silica effectively passivates the FND surface and increases the separation between the NV centers and the external interface, thereby preserving stable relaxation signals even in the presence of micromolar concentrations of gadolinium ions. Second, the intrinsic relaxation properties of the NV centers exhibit excellent temporal stability over continuous measurements extending several hours. More importantly, control experiments indicate that the observed changes in the relaxation behavior of MS-silica-FND are attributable to factors other than Gd(III) ions. Even when the concentration of Gd(DTPA)$^{2-}$ is reduced to 1 fM, a measurable increase in the relaxation rate of MS-silica-FND persists (Figure 3b). Further investigations reveal that the signals detected by MS-silica-FND relaxometry originate from hydroxyl radicals generated during water splitting. Control experiments were conducted using 1 fM Gd(NO$_3$)$_3$ (Figure 3b), showing that $T_1$ remained unchanged in the presence of nitrate ions which acting as radical quencher. Furthermore, when DMSO is used as a scavenger for hydroxyl radicals, the $T_1$ relaxation time of MS-silica-FND also remained unchanged, even with doping of 1 fM concentrations of Gd(DTPA)$^{2-}$ (Supplementary Figure 5). A more detailed comparison result in difference solutions can be seen in Supplementary Figure 6.

The products of water splitting are hydroxyl radicals, which have been verified by traditional electron paramagnetic resonance (EPR) spectroscopy. A X-band continuous-wave (CW)-EPR spectroscopy, with 5,5-dimethyl-1-pyrroline-N-oxide (DMPO) as a spin trapping agent, was employed to capture and confirm the presence of hydroxyl radicals (Figure 3c). Conventional EPR measurements have further confirmed that hydroxyl radicals are generated via water splitting on the surface of gadolinium-doped porous silica under light-free conditions. This photocatalysis-independent method holds promise for expanding application prospects in light-free environments. However, water molecules typically resist spontaneous dissociation into reactive free radicals. The relevant details are elaborated sequentially.

Unlike photocatalytic of titanium dioxide nanoparticles or ultrasonic sonochemistry methods for hydroxyl radical generation via water splitting, gadolinium-doped porous silica does not rely on external stimuli such as light or ultrasound. The mechanism, as elucidated by density functional theory (DFT) calculations, is as follows: Water molecules generally face difficulty in spontaneously dissociating into ·H and ·OH (with a water dissociation energy of $\Delta E_{dec}$ = 122.2 kcal/mol). However, dissociation occurs more readily on the surface of porous silica ($\Delta E_{dec}$ = -44.10 kcal/mol), and is further facilitated by $Gd^{3+}$ catalysis ($\Delta E_{dec}$ = -56.91 kcal/mol) (Figure 3d). Geometry optimizations for the adsorption of ·H and ·OH at various sites on the $(SiO2)_8$ cluster[30] reveal that the most stable configuration (Figure 3e) has ·OH adsorbed near the Si atom with an adsorption energy of -63.8 kcal/mol, while ·H is adsorbed at the edge O atom with an adsorption energy of -72.7 kcal/mol. Water molecules adsorb on the Si atom with an adsorption energy of -22.5 kcal/mol. These results suggest that both ·OH and ·H can stably adsorb on the surface of the clusters.

Further geometry optimization of $Gd^{3+}$ adsorption on the porous silica surface was performed to assess its catalytic potential. $Gd^{3+}$ was found to adsorb between two oxygen atoms of $(SiO_2)_8$, with an adsorption energy of -399.7 kcal/mol. Figure 3f illustrates the most stable structure for the adsorption of ·H, ·OH, and $H_2O$ on the $Gd^{3+}$-$(SiO_2)_8$ cluster, along with the corresponding adsorption energies. Specifically, ·H adsorbs at the edge O atom of porous silica with an adsorption energy of -124.9 kcal/mol, ·OH adsorbs near the Si atom with an adsorption energy of -92.8 kcal/mol, and $H_2O$ adsorbs atop the Si atom with an adsorption energy of -51.2 kcal/mol.

The conventional transition state theory was applied to determine the rate constant for the water dissociation reaction. The rate of water dissociation on gadolinium-doped porous silica is $2.2 \times 10^9$ times faster than on pure porous silica. These results demonstrate that gadolinium significantly enhances catalytic activity, accelerating the dissociation process and improving performance relative to non-gadolinium catalysis. This highlights the potential of gadolinium doping to enhance porous silica for applications in water splitting and free radical generation. For further details on the DFT calculations, see Supplementary Figure 7-10.

Hydroxyl radicals prepared by various methods are mainly used in wastewater treatment and cancer therapy (see Supplementary Tables 2). Mesoporous silica doped gadolinium (III) complex also exhibits enhanced germicidal activity and catalytic sterilization even in dark condition. The relevant antibacterial effects against *Escherichia coli* are presented in the Supplementary Figure 11. Although wastewater treatment has become a highly mature industrial technology, the process that uses water instead of hydrogen peroxide as the reactant and does not require photocatalytic hydrolysis still has broad application prospects and potential applicable scenarios. Controllable preparation and regulation will lay the foundation for expanding a wider range of applications. The following text will focus on this discussion.

**Controlled hydroxyl radical generation with real-time, in situ monitoring via MS-silica-FND**

According to simulation calculations based on conventional transition state theory, gadolinium significantly enhances catalytic activity compared to non-gadolinium systems. To improve the controllability of hydroxyl radical generation during water splitting, various concentrations of gadolinium (III) catalyst were employed, ranging from micromolar levels down to the low tens of attomolar range, and adsorbed by the mesoporous silica shell. Figure 4a illustrates the evolution of the steady-state $T_1$ relaxation rate of MS-silica-FND at varying concentrations of Gd(DTPA)$^{2-}$. The data show that the $\Delta\Gamma$ decreases as the concentration decreases to 10 fM, with a rapid drop observed between 100 aM to 10 aM. The calculated concentrations of generated hydroxyl and hydrogen radicals, as well as the correlation between radical concentrations and the corresponding gadolinium (III) catalyst concentrations, are shown in Figure 4b. This correlation is based on the quenching of the $T_1$ relaxation time of MS-silica-FND sensors and the extrapolation of the known contribution of paramagnetic ions from the detection results. Hydroxyl radical generation was observed to occur at a controllable steady-state level, ranging from low concentrations (~100 mM) up to few-mol. As previously described, changes in $T_1$ relaxation time depend on the distance between the magnetic signal from the radicals and the NV centers. Therefore, variations in the shape of the nanodiamonds and the distribution of NV centers within them must be considered. Additional calibration experiments and calculations are detailed in the Supplementary Figure 5-6 and Supplementary Figure 12-18, respectively. Under the conditions shown in Figure 4b, this method yields radical concentrations ranging from micromolar to molar per liter, corresponding to gadolinium (III) catalyst concentrations in the range of attomolar to picomolar per liter. A comparison between the model of gadolinium (III) ion contributions and experimental results, which include both gadolinium (III) ions and generated free radicals, shows that, at low concentrations of catalytic paramagnetic ions, the reduction in $T_1$ is primarily driven by high-frequency magnetic noise from free radicals (Supplementary Figure 17). The specific concentrations of hydrogen radicals can vary depending on the tailored catalytic conditions and the core-shell structure of MS-silica-FND.

NV radical sensing, used to monitor radical generation, shows good stability, and the generation of active free radicals within the mesoporous silica layer with gadolinium (III) demonstrates high efficiency and stability. As part of the nanoscale catalysis evaluation, the homogeneity of catalytic activity among individual particles was examined to determine whether free radicals were generated by only a few nanoparticles, particularly at low gadolinium (III) doping concentrations in the attomolar per liter range. The $T_1$ relaxation rate changes of 177 randomly selected MS-silica-FND particles were in situ monitored and analyzed individually by switching the solution from Milli-Q water to a 100 aM $Gd(DTPA)^{2-}$ solution (results shown in Figure 4c). The histogram of the relaxation rate changes indicates that most nanodiamonds exhibited a positive change. These results suggest that gadolinium (III) was precisely doped onto the mesoporous silica shell with controlled concentration. In conclusion, through real-time, in situ monitoring using MS-silica-FND, the system enables long-term stable radical production with precisely controllable concentrations.

**Discussion and Conclusion**

This work demonstrates a significant advancement in the field of quantum sensing and reactive species monitoring by integrating targeted material design with the unique capabilities of nitrogen-vacancy (NV) centers in fluorescent nanodiamonds (FNDs). The core achievement lies in the rational design and synthesis of a double-layered silica architecture on FNDs (MS-silica-FND), which simultaneously addresses two fundamental challenges in FND-based paramagnetic sensing: protecting the quantum sensor from environmental noise and creating a functional interface for capturing transient analytes.

The dense inner silica shell successfully passivates the FND surface, shielding near-surface NV centers from magnetic noise and preserving their quantum coherence and relaxation properties, as evidenced by the stable $T_1$ signals of silica-FND in control experiments. This foundational stability is critical for achieving reliable, high-sensitivity detection. Concurrently, the outer mesoporous silica layer serves a dual purpose: it acts as a high-surface-area scaffold for the efficient adsorption and stabilization of short-lived hydroxyl radicals, and it provides a versatile platform for doping with metal ion catalysts. Among the various available metal ion-doped catalysts, $Gd(DTPA)^{2-}$ was selected due to its status as an FDA-approved pharmaceutical agent.

Concurrently, leveraging the biocompatibility of nanodiamonds and the low toxicity of silica, the entire nanosystem exhibits substantial potential for applications in biological environments.

The mechanism of light-free, catalytic water splitting on the gadolinium-doped porous silica surface represents a key innovation. DFT calculations convincingly show that the porous silica surface, especially when doped with gadolinium (III), dramatically lowers the energy barrier for water dissociation. This enables the continuous and on-demand generation of hydroxyl radicals without relying on external energy inputs like light or ultrasound, or on chemical precursors like hydrogen peroxide. This generating with sensing paradigm distinguishes our platform from conventional detection methods, which often require the external addition of the target species or its precursors.

The real-time, in situ monitoring capability of the MS-silica-FND platform is powerfully validated through $T_1$ relaxometry. The sensor's response is directly proportional to the local concentration of paramagnetic species (·OH and ·H), allowing for the quantitative tracking of radical generation kinetics. The ability to tune the steady-state radical flux by varying the concentration of the Gd(III) catalyst adsorbed in the mesoporous shell highlights the exceptional controllability of this system. Furthermore, single-particle analysis confirms the homogeneity of catalytic activity across the nanoparticle ensemble, underscoring the robustness and reproducibility of the synthesis and functionalization process.

This nanoplatform successfully bridges the gap between the need for stable, spin quantum sensors and the requirement for dynamic, reactive chemical interfaces. It transforms the FND from a passive sensor of external magnetic fields into an active component in a catalytic sensing system. This work establishes a robust and versatile framework for the persistent monitoring of highly reactive, transient species in aqueous environments. It moves beyond mere detection to create a closed-loop system where the generation of the target species is intrinsically linked to and regulated by the sensing mechanism. The principles demonstrated here—rational core-shell design, catalytic integration, and quantum relaxometry readout—conducive to the development of intelligent quantum sensors. These sensors hold great promise for advancing fundamental studies and applications in fields such as real-time monitoring of redox biology, catalytic reaction dynamics, targeted radical-based therapies, and intelligent chemical manufacturing, where precise, in situ control over reactive intermediates is paramount.

**Materials and Methods**

**Materials**

The nanodiamonds we used are 40 nm carboxylated ND (Adamas NDNV40nmHi10ml), 40 nm hydroxylated ND (Adamas, NDNV40nmHiOH2mL) and 40 nm pristine ND (Adamas ND-HPHT-NV40nm). For the synthesis of the silica-ND nanoparticles, we used those chemicals below: Tetraethyl orthosilicate (TEOS, Sigma-Aldrich 333859-25ML), Ammonium hydroxide solution (28.0%-30.0% $NH_4OH$, Sigma-Aldrich 221228-25ML-A), Ethanol absolute (EtOH, Sinopharm 10009218). For the synthesis of the MS-silica-ND nanoparticles, we used those chemicals below in addition to the above: Ammonium-$^{15}$N nitrate (Abundance: 10atom% $^{15}NH_4NO_3$, Macklin A801378-25g), Hexadecyltrimethylammonium bromide (CTAB, Sigma-Aldrich H6269-100G). The paramagnetic labels to attach are gadopentetate dimeglumine ($Gd(DTPA)^{2-}$, MedChemExpress MCE-HY-107353) and pentetic acid (DTPA, MedChemExpress MCE-HY-B1335). The Milli-Q water having a resistivity of 18.2 $M\Omega \cdot cm$ is used to dilute $Gd(DTPA)^{2-}$ to different concentrations. Dimethyl sulfoxide (DMSO, Sinopharm 30072418) is used for quenching hydroxyl radicals in control experiment. The spin trapping agent to capture hydroxyl radicals for EPR measurement is 5,5-dimethyl-1-pyrroline-N-oxide (DMPO, Dojindo D048). All the reagents were analytically pure (AR) and without further purification.

**DFT calculations.**

The DFT calculations were performed with the B3LYP method for describing intermolecular interactions using the Gaussian-16 package. The basis set 6-31+G(d,p) was used for hydrogen atoms and oxygen atoms. The SDD basis set was used for the Gd ion and Si atoms[31]. The geometry optimizations were performed via the GDIIS algorithm[32] with the convergence criteria of a maximum step size of 0.0018 au and a root mean square (RMS) force of 0.0003 au.

**Synthesis.**

The synthesis of nanoparticles mainly has two parts, silica-FND and mesoporous silica-FND (MS-silica-FND) respectively. For the synthesis of silica-FND, a homogeneous dispersion was prepared by mixing 40 nm carboxylated FND with ethanol (EtOH) and ammonium hydroxide ($NH_4OH$) at a volume ratio of 1 mL : 15 mL : 150 μL. Subsequently, tetraethyl orthosilicate (TEOS) was introduced dropwise at a ratio of 1 mL FND suspension to 5 μL TEOS under mild stirring to ensure controlled hydrolysis and condensation, thereby enabling the formation of a uniform silica coating on the FND surface. After stirring for 10 h under controlled conditions (1000 rpm, 25 °C), the nanoparticles were collected by centrifugation at 3600 rpm for 25 min and subsequently redispersed in EtOH. The synthesis of silica-FND was completed following two consecutive cycles of this purification and surface modification process.

For the synthesis of MS-silica-FND, EtOH, Milli-Q water, and $NH_4OH$ were mixed in a volume ratio of 1.45 mL : 2.15 mL : 20 µL per 1 mL of silica-FND suspension. The corresponding volume of silica-FND was then added under ultrasonication, followed by an additional 30 min of ultrasonication to ensure uniform dispersion. Subsequently, cetyltrimethylammonium bromide (CTAB) was introduced at a ratio of 20 mg CTAB : 330 µL Milli-Q water : 150 µL EtOH per 1 mL silica-FND, and the mixture was further sonicated for 30 min. Thereafter, TEOS was added dropwise at a ratio of 10 µL TEOS per 1 mL silica-FND, and the resulting solution was stirred overnight at 1000 rpm and 25 °C to promote mesoporous silica shell growth. The product was collected by centrifugation at 10,000 rpm for 5 min, repeated twice: during the first cycle, the precipitate was redispersed in ethanol; during the second, in an ammonium nitrate solution (330 mg $NH_4NO_3$ in 100 mL of 95% EtOH). After each centrifugation step, the purified particles were shaken in a shaker at 1500 rpm and 60 °C for 1 h to facilitate template removal. Finally, following two complete iterations of centrifugation and redispersion in EtOH, the MS-silica-FND was obtained.

**Transmission Electron Microscope (TEM).**

TEM images were taken by using a JEM-2100 plus microscope. The samples used for TEM were prepared by depositing synthesized FNDs dispersed in EtOH on carbon-coated copper girds and dried by an infrared lamp.

**Dynamical Light Scattering (DLS).**

The hydrodynamic diameter distribution and Zeta potential were taken by using a particle analyzer (Brookhaven NanoBrook Omni). The hydrodynamic diameter was measured by diluting synthesized FNDs in Milli-Q to 0.01 mg/mL with a 90º angle at room temperature. The Zeta potential was measured at room temperature.

**Electron Paramagnetic Resonance (EPR).**

The generation of hydroxyl radicals was detected using X-band continuous-wave electron paramagnetic resonance (CW-EPR) spectroscopy (CIQTEK, EPR200-Plus). To enable sufficient signal accumulation for reliable detection, 5,5-dimethyl-1-pyrroline N-oxide (DMPO, Dojindo, D048) was employed as a spin trapping agent to capture transient hydroxyl radicals and stabilize them for measurement. Immediately after the addition of 20 µL DMPO to 180 µL of Gd(DTPA)$^{2-}$-doped mesoporous silica nanoparticle suspension, the EPR spectrum was recorded.

**Optical setup for $T_1$ relaxation time measurement.**

A home-built wide-field fluorescence microscope was employed to optically detect the nanoparticle samples. An inverted Olympus LUCPLFLN60X objective (numerical aperture, NA = 0.7) was used for excitation and emission collection. The sample was mounted on a custom-designed adapter and positioned using a piezoelectric scanner (CoreMorrow N12.XYZ100S) for fine positioning, and a motorized linear stage (Thorlabs Z825B) for coarse adjustment. Optical excitation was provided by a 532 nm laser (Lighthouse Sprout-D-5W), intensity-modulated via an acousto-optic modulator (Isomet M1133-aQ80L-2.0). The resulting fluorescent signal was collected through a 650 nm long-pass filter (Thorlabs FELH0650) and directed to either an avalanche photodiode detector (Thorlabs APD410A/M) for ensemble-level measurements or a scientific CMOS camera (Dhyana 95 V2) for single-particle detection, with optical path selection controlled by a motorized filter flip mount (Thorlabs MFF101/M). Pulse sequences were generated and synchronized using a 500 MHz digital pulse generator (SpinCore Technologies PulseBlasterESR-PRO).

**Sample preparation for $T_1$ relaxation time measurement.**

The synthesized nanoparticle in EtOH with a density of around 1 mg/mL was dropped on the cover glass (24 mm×50 mm, VWR 16004-098) treated with oxygen plasma gas and dried naturally. After that, the cover glass was washed with a pipette and dried naturally several times to obtain an almost single layer sample (see the SEM result Supplementary Figure 4). A gasket with a depth of 0.12 mm and a well of 13 mm diameter (Thermo Fisher Scientific S24735) was attached to the cover glass with sample in the well to contain 75 µL liquid (Milli-Q or Gd(DTPA)$^{2-}$ solution with concentration changed). Another cover glass (22 mm×22 mm, VWR 16004-094) treated with oxygen plasma gas was used for the cover to avoid evaporation of the solution.

**Acknowledgments**

We thank Kangwei Xia, Ya Wang and Rui He for helpful discussion. This work was supported by the National Natural Science Foundation of China (Grant No. T2125011, No.12105166), the CAS (YSBR-068), Innovation Program for Quantum Science and Technology (Grant No. 2021ZD0302200, 2021ZD0303204), New Cornerstone Science Foundation through the XPLORER PRIZE, and the Fundamental Research Funds for the Central Universities (Grant No. WK 3540000010), the National Key R&D Program of China (Grant No. 2024YFB3212600, 2024YFE0206900), the Natural Science Foundation of Qinghai Province of China (2023-ZJ-940J), Kunlun Talents-High end Innovation and Entrepreneurship Talent Plan of Qinghai Province.

**Author Contributions:** F.S. supervised the project; F.S. and J.S. proposed the idea and designed the experiments. J.S. and Z.K. performed the experiments; Z.K. and F.K. performed NV related calculations; G.S., X.L. and J.L. carried on the DFT calculations; J.S., J.Y. and Z.Z. performed the sterilization experiments; J.H.S. suggested on the EPR analysis; Z.K., J.S., F.K. and Z.W. built the setup. J.S., Z.K. and F.S. wrote the manuscript. The manuscript was written through contributions of all authors. All authors discussed and analyzed the data

**References**

1. H.-C. Chang, W. W. W. Hsiao, M. C. Su, Fluorescent Nanodiamonds (Wiley, 2018)
2. A. Gruber, A. Drabenstedt, C. Tietz, L. Fleury, J. Wrachtrup, C. V. Borczyskowski, Scanning confocal optical microscopy and magnetic resonance on single defect centers. *Science* 276, 2012–2014 (1997).
3. R. Schirhagl, K. Chang, M. Loretz, C. L. Degen, Nitrogen-vacancy centers in diamond: Nanoscale sensors for physics and biology. *Annu. Rev. Phys. Chem.* 65, 83–105 (2014)
4. Y. Wu, T. Weil, Recent developments of nanodiamond quantum sensors for biological applications. *Advanced Science* 9, 2200059 (2022).
5. J. Su, Z. Kong, L. Zeng, et al., Fluorescent nanodiamonds for quantum sensing in biology. *WIREs Nanomed Nanobiotechnol* 17, e70012 (2025).
6. J. Holzgrafe, Q. Gu, J. Beitner, D. M. Kara, H. S. Knowles, M. Atatüre, Nanoscale NMR spectroscopy using nanodiamond quantum sensors. *Phys. Rev. Appl.* 13, 044004 (2020).
7. Z. Qin, Z. Wang, F. Kong, et al., In situ electron paramagnetic resonance spectroscopy using single nanodiamond sensors. *Nat. Commun.* 14, 6278 (2023).

8. G. Kucsko, P. C. Maurer, N. Y. Yao, et al., Nanometre-scale thermometry in a living cell. *Nature* 500, 54–58 (2013).
9. C.-F. Liu, W. H. Leong, K. Xia, et al., Ultra-sensitive hybrid diamond nanothermometer. *Natl. Sci. Rev.* 8, nwaa194 (2021).
10. A. Sigaeva, N. Norouzi, R. Schirhagl, Intracellular relaxometry, challenges, and future directions. *ACS Cent. Sci.* 8, 1484–1489 (2022).
11. J.-P. Tetienne, T. Hingant, L. Rondin, A. Cavaillès, L. Mayer, G. Dantelle, T. Gacoin, J. Wrachtrup, J.-F. Roch, V. Jacques, Spin relaxometry of single nitrogen-vacancy defects in diamond nanocrystals for magnetic noise sensing. *Phys. Rev. B* 87, 235436 (2013).
12. Y. Wu, P. Balasubramanian, Z. Wang, J. A. Coelho, M. Pršlja, R. Siebert, M. B. Plenio, F. Jelezko, T. Weil, Detection of few hydrogen peroxide molecules using self-reporting fluorescent nanodiamond quantum sensors. *J. Am. Chem. Soc.* 144, 12642–12651 (2022).
13. J. Vavra, I. Rehor, T. Rendler, et al., Supported lipid bilayers on fluorescent nanodiamonds: A structurally defined and versatile coating for bioapplications. *Adv. Funct. Mater.* 28, 1803406 (2018).
14. S. Iyer, C. Yao, O. Lazorik, M. S. Bin Kashem, P. Wang, G. Glenn, M. Mohs, Y. Shi, M. Mansour, E. Henriksen, K. Murch, S. Mukherji, C. Zu, Optically trapped nanodiamond relaxometric detection of nanomolar paramagnetic spins in aqueous environments. Phys. Rev. Appl. 22, 064076 (2024).
15. T. Rendler, J. Neburkova, O. Zemek, et al., Optical imaging of localized chemical events using programmable diamond quantum nanosensors. *Nat. Commun.* 8, 14701 (2017).
16. J. Barton, M. Gulka, J. Tarabek, et al., Nanoscale dynamic readout of a chemical redox process using radicals coupled with nitrogen-vacancy centers in nanodiamonds. *ACS Nano* 14, 12938–12950 (2020).
17. K. Żamojć, M. Zdrowowicz, D. Jacewicz, D. Wyrzykowski, L. Chmurzyński, Fluorescent and luminescent probes for monitoring hydroxyl radical under biological conditions. Crit. *Rev. Anal. Chem.* 46, 160–169 (2016).
18. F. Perona Martínez, A. C. Nusantara, M. Chipaux, S. K. Padamati, R. Schirhagl, Nanodiamond relaxometry-based detection of free-radical species when produced in chemical reactions in biologically relevant conditions. *ACS Sens.* 5, 3862–3869 (2020).
19. J. Barton, M. Gulka, J. Tarabek, et al., Nanoscale dynamic readout of a chemical redox process using radicals coupled with nitrogen-vacancy centers in nanodiamonds. *ACS Nano* 14, 12938–12950 (2020).
20. L. Nie, A. C. Nusantara, V. G. Damle, et al., Quantum monitoring of cellular metabolic activities in single mitochondria. *Sci. Adv.* 7, eabf0573 (2021).

21. J. Su, L. Zeng, P. Chen, Z. Kong, F. Shi, et al., Subcellular metabolic tracking using fluorescent nanodiamonds relaxometry. *Adv. Funct. Mater.* e27416 (2025)
22. C. C. Winterbourn, Reconciling the chemistry and biology of reactive oxygen species. *Nat. Chem. Biol.* 4, 278–286 (2008)
23. B. D'Autréaux, M. B. Toledano, ROS as signalling molecules: Mechanisms that generate specificity in ROS homeostasis. *Nat. Rev. Mol. Cell Biol.* 8, 813–824 (2007).
24. U. Zvi, D. R. Candido, A. M. Weiss, A. R. Jones, L. Chen, I. Golovina, X. Yu, S. Wang, D. V. Talapin, M. E. Flatté, A. P. Esser-Kahn, & P. C. Maurer, Engineering spin coherence in core-shell diamond nanocrystals. *Proc. Natl. Acad. Sci.* 122, e2422542122 (2025).
25. M. Barzegaramiriolya, E. S. Grant, T. Ralph, Y. Li, G. Thalassinos, A. Tadich, L. Thomsen, T. Ohshima, H. Abe, N. Dontschuk, A. Stacey, P. Mulvaney, L. T. Hall, P. Reineck, D. A. Simpson, Functionalized fluorescent nanodiamonds with millisecond spin relaxation times. *ACS Nano* 19, 36884–36895 (2025).
26. W. Stöber, A. Fink, E. Bohn, Controlled growth of monodisperse silica spheres in the micron size range. *J. Colloid Interface Sci.* 26, 62–69 (1968).
27. E. von Haartman, H. Jiang, A. A. Khomich, J. Zhang, S. A. Burikov, T. A. Dolenko, J. Ruokolainen, H. Gu, O. A. Shenderova, I. I. Vlasov, J. M. Rosenholm, Core–shell designs of photoluminescent nanodiamonds with porous silica coatings for bioimaging and drug delivery I: Fabrication. *J. Mater. Chem. B* 1, 2358–2366 (2013).
28. A. O. Sushkov, N. Chisholm, I. Lovchinsky, M. Kubo, P. K. Lo, S. D. Bennett, D. Hunger, A. Akimov, R. L. Walsworth, H. Park, M. D. Lukin, All-optical sensing of a single-molecule electron spin. *Nano Lett.* 14, 6443–6448 (2014).
29. P. Reineck, L. F. Trindade, J. Havlik, J. Stursa, A. Heffernan, A. Elbourne, M. Capelli, P. Cigler, D. A. Simpson, B. C. Gibson, Not all fluorescent nanodiamonds are created equal: A comparative study. *Part. Part. Syst. Charact.* 36, 1900009 (2019).
30. E. Flikkema, S. T. Bromley, Dedicated global optimization search for ground state silica nanoclusters: $(SiO_2)$ N (N = 6–12). *J. Phys. Chem. B* 108, 9638–9645 (2004).
31. C. Paduani, DFT study of gadolinium aluminohydrides and aluminofluorides. *Chem. Phys.* 417, 1–7 (2013).
32. X. Li, M. J. Frisch, Energy-represented direct inversion in the iterative subspace within a hybrid geometry optimization method. *J. Chem. Theory Comput.* 2, 835–839 (2006)

**Figures and Tables**

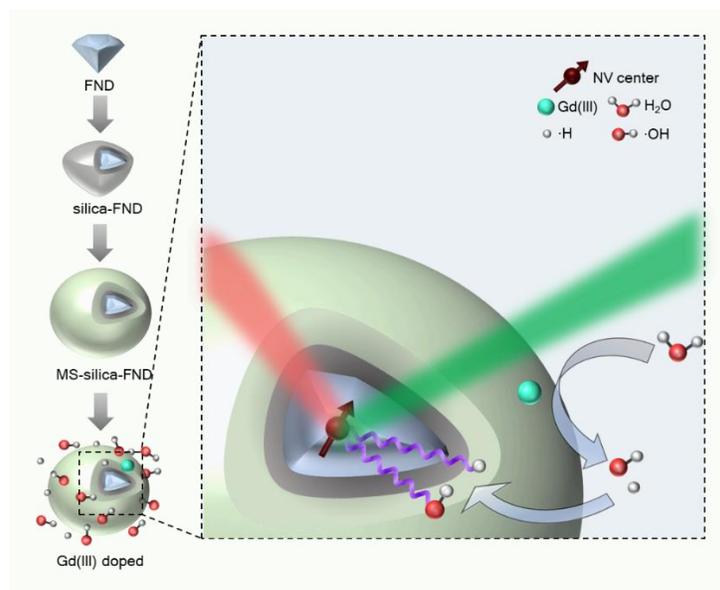

**Figure 1. Design strategy of double-layered silica-engineered FNDs for real time monitoring hydroxyl radical generation via water splitting.** Schematic illustration of gadolinium (III) doped MS-silica-FND synthesis and apply to hydroxyl radical generation through light-free catalytic water splitting with paramagnetic sensing of FNDs with NV spin.

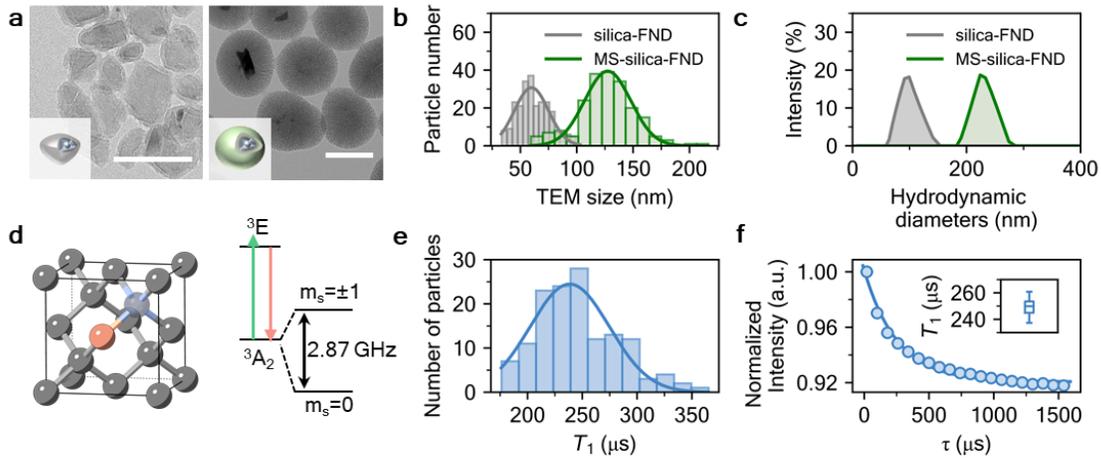

**Figure 2. Characterization of core-shell functionalized FNDs and quantum relaxation properties.** (a) The TEM images of silica-FND (left) and MS-silica-FND (right), respectively. Scale bar: 100nm. (b) Histogram of the particle size for silica-FND (gray) and MS-silica-FND (green), with the fitting line representing the Gaussian fit of the distribution, indicates a mean size of 67.0 nm and 127.3 nm in diameter, respectively. (c) DLS characterization of silica-FND (gray line) and MS-silica-FND (green line). (d) Schematic diagram of the lattice structure (left) and energy level structure (right) of NV center. The $T_1$ relaxation measurement utilizes the sublevel of the ground state to sense the high frequency magnetic noise around ~GHz level. (e) Histogram of single particle relaxation time of 40 nm-size-ND and the red line is a Gaussian fitting of the histogram yielding mean $T_1$ = 238.5 μs. (h) Spin relaxation data for ensemble particles, recorded using a PD detector. Blue line: Fitting to equation which yields $T_1$ = 255.2 μs. Inset: Boxplot of ensemble relaxation measurement (N=7).

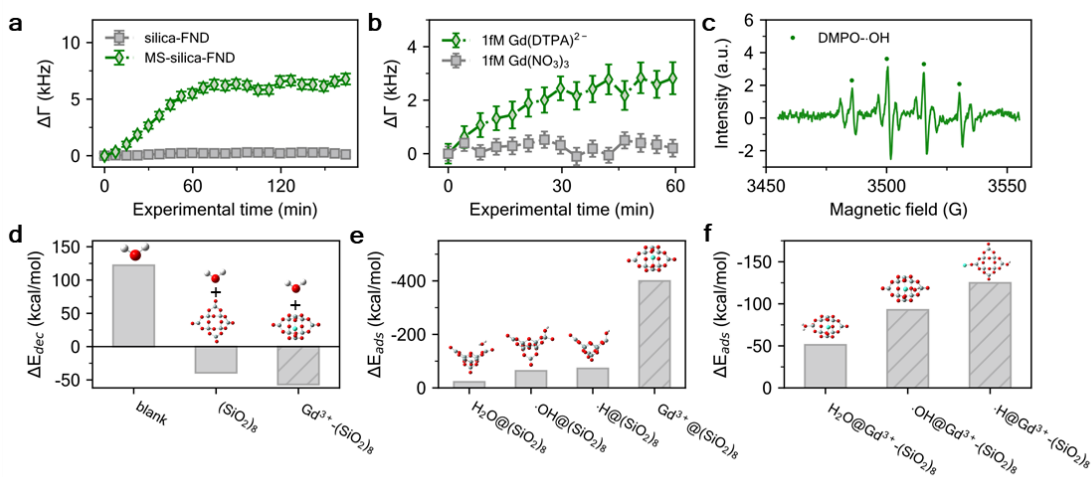

**Figure 3. FND relaxometry observation and modeling of hydroxyl radical generation.** (a) Time profiles of changed relaxation rates for silica-FND (gray square point) and MS-silica-FND (green diamond point) in 1 μM Gd(DTPA)$^{2-}$ solution. The error bars represent the fitting error. (b) Time profiles of changed relaxation for MS-silica-FND in 1 fM Gd(DTPA)$^{2-}$ solution (green diamond point) and 1 fM Gd(NO$_3$)$_3$ (gray square point). The error bars represent the fitting error. (c) CW-EPR spectrum of ·OH with spin trapping agents DMPO. (d) Calculated results of water dissociation energy ΔE$_{dec}$ in the case of dissociate spontaneously, on the surface of porous silica, and in the presence of Gd$^{3+}$ catalysis on the surface of porous silica, respectively. (b) Geometry optimizations of H$_2$O, ·OH, ·H, and Gd$^{3+}$ adsorption on the porous silica and corresponding adsorption energies ΔE$_{ads}$. (c) Geometry optimizations of H$_2$O, ·OH, and ·H adsorption on the Gd$^{3+}$ doped porous silica, and corresponding adsorption energies ΔE$_{ads}$ respectively.

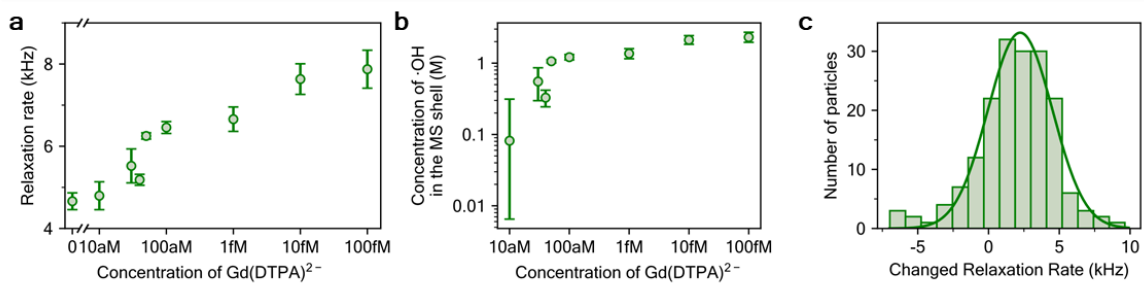

**Figure 4. Controlled hydroxyl radical generation with real-time, in situ monitoring via MS-silica-FND.** (a) Evolution of the steady-state $T_1$ relaxation rate of MS-silica-FND at varying concentrations (from 10 aM to 100 fM) of Gd(DTPA)$^{2-}$ in solution. The error bars represent the standard deviation within three groups for each point. (b) The simulation concentrations of radicals in the mesoporous silica shell and correlation between radical concentrations and the corresponding Gd(DTPA)$^{2-}$ catalyst concentrations. The error bars represent the standard deviation generated by the random position of NV center within FND in the Monte Carlo simulation. (c) Histogram of changing in 177 single-particle $T_1$ relaxation rates for MS-silica-FND when the solution is switched from Milli-Q to 100 aM Gd(DTPA)$^{2-}$.